\documentclass[%
 reprint,
superscriptaddress,
nofootinbib,
 amsmath,amssymb,
 aps,
 prx,
]{revtex4-2}

\usepackage{graphicx}% Include figure files
\usepackage{dcolumn}% Align table columns on decimal point
\usepackage{bm}% bold math
\usepackage{miller}
\usepackage{setspace}
\usepackage{siunitx}
\usepackage{nicefrac}
\usepackage{enumerate}
\usepackage{multirow}
\usepackage[disable]{todonotes}
\usepackage{tabularx}
\usepackage{makecell} 
\usepackage{tablefootnote}
\usepackage{amssymb}
\usepackage{amsmath}
\usepackage{comment}
\usepackage{hyperref}  % Required for cross-file hyperlinks
\usepackage{xr}  % cross-referencing
\usepackage{soul}
\usepackage{float}
\floatstyle{plaintop}
\restylefloat{table}
\usepackage{enumitem}
\begin{document}

\preprint{prl}

%\title{The vibrational dynamics of a single layer of MoS\textsubscript{2}}
%\title{Acoustic phonon dispersion in a single layer of MoS\textsubscript{2}}
%\title{Defect-modified acoustic phonon dispersion in a monolayer semiconductor}
\title{Defect-modified acoustic phonons in a single layer of MoS\textsubscript{2}}
 
\author{Aleksandar Radi\'{c}}\email{ar2071@cam.ac.uk}
\affiliation{Cavendish Laboratory, Department of Physics, University of Cambridge, JJ Thomson Ave, Cambridge, United Kingdom}%

\author{Boyao Liu}
\affiliation{Cavendish Laboratory, Department of Physics, University of Cambridge, JJ Thomson Ave, Cambridge, United Kingdom}%

\author{Akshay Rao}
\affiliation{Cavendish Laboratory, Department of Physics, University of Cambridge, JJ Thomson Ave, Cambridge, United Kingdom}%

\author{Sam M Lambrick}%\email{sml59@cantab.ac.uk}
\affiliation{Cavendish Laboratory, Department of Physics, University of Cambridge, JJ Thomson Ave, Cambridge, United Kingdom}%
\affiliation{ISIS Facility, Rutherford Appleton Laboratory, Chilton, Didcot, Oxfordshire OX11 0QX, United Kingdom}

\date{\today}% It is always \today, today,

\begin{abstract}
The thermal, mechanical, and electronic performance of atomically thin semiconductors is governed by their low-energy phonons, yet the impact of atomic-scale disorder on these modes remains poorly understood. Here, we report the first measurement of acoustic phonon dispersions in a quasi-freestanding monolayer semiconductor (MoS\textsubscript{2}), using helium-3 spin-echo spectroscopy. We identify a defect-driven regime change at a critical wavevector, $q_c$, marking the breakdown of continuum elastic behavior. At this length scale, the flexural mode transitions from continuum bending to defect-pinned standing waves, while the hybridized Rayleigh wave becomes vibrationally disordered in its dispersion and linewidth. We observe multiple defect-induced Van Hove singularities deep within the Brillouin zone and strongly suppressed acoustic group velocities, providing direct experimental evidence that four-phonon processes drive thermal transport in mono- and few-layer MoS\textsubscript{2}. These results offer a microscopic explanation for the anomalously low thermal conductivity widely observed in transition-metal dichalcogenides and demonstrate how atomic-scale disorder dictates energy flow in two-dimensional materials.
\end{abstract}

\maketitle
%\phantom{s}
%\clearpage

\section{Introduction}

Atomically thin semiconductors have emerged as a versatile platform for next-generation optoelectronics, flexible electronics, and quantum technologies. Transition metal dichalcogenides (TMDs) are of specific interest  due to their unique optical and electronic properties that are highly tunable via thickness\cite{mak_atomically_2010,Kuc2011,Ma2014}, defect engineering\cite{zhu_room-temperature_2023}, and surface functionalization\cite{Nie2020,Bretscher2021}. However, their integration into high-performance devices faces a fundamental bottleneck: anomalously poor thermal transport. Suspended monolayer MoS\textsubscript{2} exhibits experimental thermal conductivities\cite{Yan2014,Sahoo2013,Jo2014} more than an order of magnitude lower than both theory\cite{Wu2021,Ding2015,Peng2016,Li2013,Cai2014,Xu2016} and monolayer graphene\cite{Han2023ThermalDiamond}. The broad range of predicted values, the closest of which is a factor of two larger than experimental values, indicates a lack of understanding of vibrational dynamics in 2D systems. This dramatic suppression of heat dissipation is typically attributed to `surface disorder'\cite{Yarali2017EffectsDeposition,Polanco2020Defect-limitedMoS2}, limiting device performance and constraining the power density achievable in TMD-based technologies. Understanding thermal transport in 2D materials with non-negligible defect densities, like TMDs, requires insight into the limitations of models currently used to describe these lattice vibrations, and critically the point at which disorder begins to dominate.

The origins of the thermal bottleneck lie in the low-energy acoustic phonons that dominate heat transport. In two-dimensional (2D) systems, these modes are expected to follow continuum elastic mechanics at long wavelengths, with deviations emerging only near the Brillouin zone (BZ) edge. This continuum description, however, neglects atomic-scale disorder in the form of vacancies, substitutional defects, and grain boundaries that are unavoidable in real materials. While there exists extensive literature on the electronic structure of TMDs, their vibrational dynamics, particularly in the mono- and few-layer limit where confinement amplifies the role of defects, remain largely unexplored. Recent first-principles calculations suggest that four-phonon scattering processes, defect-phonon interactions, and localized vibrational states may dominate thermal transport\cite{Peng2016BeyondMoS2,Polanco2020Defect-limitedMoS2,Chaudhuri2024UnderstandingMoS2}, but direct experimental validation has been lacking. Understanding how and where continuum descriptions break down due to disorder is therefore central to explaining the anomalously low thermal conductivity of mono- and few-layer TMDs.

Here we probe low-energy acoustic phonons in monolayer MoS\textsubscript{2}, the prototypical TMD, and reveal how atomic-scale disorder reshapes their vibrational dynamics well before the BZ edge is reached. By probing the topmost layer of bulk MoS\textsubscript{2}, which behaves as a quasi-freestanding monolayer, we obtain direct experimental insight into disorder-controlled phonon behavior in a two-dimensional semiconductor.

We identify a defect-controlled length scale at which continuum descriptions of acoustic phonons break down in monolayer MoS\textsubscript{2}.  This disorder-driven cross-over produces multiple Van Hove singularities away from the BZ edge, suppression of phonon group velocities, and limits the phonons’ lifetimes – all factors that contribute significantly to thermal transport. These results establish a microscopic connection between atomic-scale disorder and macroscopic thermal transport in two-dimensional semiconductors, clarifying why continuum elastic models systematically fail in materials with non-negligible defect densities.

\subsection{Helium-3 spin-echo spectroscopy}

\begin{figure}[H]
    \centering
    \includegraphics[width=85mm]{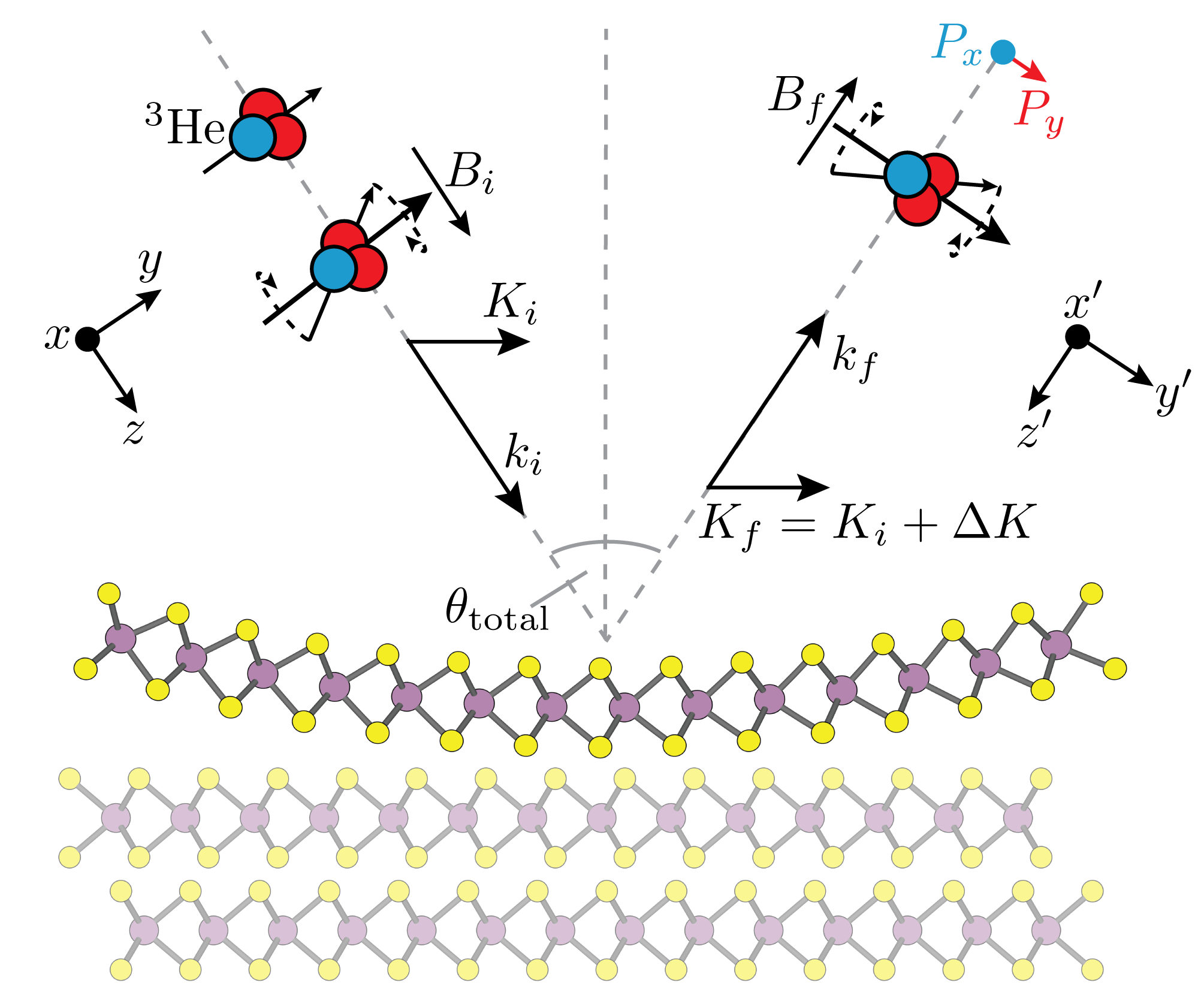}
    \caption{Schematic of helium-3 spin-echo (HeSE) spectroscopy. The topmost layer of bulk MoS\textsubscript{2} vibrates as a quasi-freestanding monolayer due to weak interlayer van der Waals (vdW) binding. The spins of the incident \textsuperscript{3}He are polarized and precession induced by a solenoidal magnetic field ($B_i$). During scattering the atoms can exchange energy with the surface, exciting or destroying surface phonons. The final energy distribution is measured in the time domain \emph{via} the atoms' perpendicular spin polarizations ($P_x,P_y$).}
    \label{fig:spineecho}
\end{figure}

Helium-3 spin-echo spectroscopy (HeSE),  the core principle of which is shown in Figure \ref{fig:spineecho}, is a surface-sensitive scattering technique that measures atomic-scale dynamics by encoding energy and momentum exchange between a beam of spin-polarized helium atoms and a surface into the Larmor precession of the helium nuclear spins\cite{Jardine2004,alexandrowicz_helium_2007}. Following Figure \ref{fig:spineecho}, \textsuperscript{3}He atoms are first spin-polarized using a permanent-magnet hexapole and then pass through a solenoid producing a magnetic field $B_i$ along the $z$-axis, which induces Larmor precession of the nuclear spins.

Upon scattering from the surface, dynamical processes such as adsorbate diffusion or lattice vibrations (phonons) lead to small changes in the atomic trajectories and velocities, thereby perturbing the accumulated spin phase. A second solenoidal field, $B_f$, is then applied after scattering to exactly reverse the precession induced by $B_i$, such that spins would refocus entirely in the absence of surface dynamics. Finally, a magnetic hexapole analyzer and ionizing detector convert the resulting spin polarization into an intensity signal, with any loss of polarization directly reflecting the time-dependent motion of atoms at the surface. HeSE records the spin polarization in the time domain as a function of the applied magnetic field integral, from which inelastic excitations can be identified.

HeSE probes only the valence electronic wavevectors of the material's topmost atoms due to the neutrally charged and millielectronvolt ($\sim\SI{8}{\milli\electronvolt}$) probe atom energies. In addition, HeSE is free from optical selection rules, enabling the detection of all vibrational modes\cite{Benedek2018}, including low-energy acoustic phonons that are often inaccessible to conventional spectroscopies. Converting from the measured time-domain of the scattered spin polarization to the energy-domain yields the phonon dispersion and linewidth, allowing simultaneous access to mode energies, lifetimes, and damping mechanisms\cite{liu_distinguishing_2024,liu_experimental_2024}.

For a fixed scattering geometry, such as shown in Figure \ref{fig:spineecho} where $\theta_{\mathrm{total}}=\SI{44.4}{\degree}$, the spin polarization is measured as a function of magnetic field integral and Fourier transformed to obtain the energy-resolved spectrum. Inelastic features are identified as phonon creation or annihilation peaks and fitted to extract mode energies and linewidths. By varying the crystal tilt angle, the momentum transfer parallel to the surface is systematically changed, allowing the phonon energies to be mapped as a function of wavevector. Repeating this procedure across momentum space yields the phonon dispersion curves, while linewidth analysis provides access to phonon lifetimes. %Full details of the fitting procedures and data processing are provided in the Appendix.

The sample measured was a synthetic flux grown single crystal purchased from 2D semiconductors Ltd.\cite{2DSemiconductors_2024}. The crystal was mechanically exfoliated \textit{ex situ} multiple times prior to transfer into the ultrahigh-vacuum scattering chamber of the HeSE spectrometer. The sample was heated to $\SI{240}{\degreeCelsius}$ for $\SI{1}{\hour}$ to clean the surface of physisorbed contaminants. The sample temperature is monitored by a K-type thermocouple and heated by a tungsten filament from the back side. The temperature of the sample was $\SI{120}{\degreeCelsius}$ and the beam energy $\SI{8.1}{\milli\electronvolt}$ for all measurements. The \textsuperscript{3}He beam was at $\SI{15}{\bar}$ for all measurements\cite{Kelsall2025}, and the pressure of the scattering chamber was $\sim 1 \times 10^{-9}\mathrm{mbar}$. A custom high sensitivity solenoidal detector\cite{chisnall_high_2012,bergin_low-energy_2021} is used to measure the \textsuperscript{3}He signal. All data were collected with a tilt angle of $\alpha=\ang{135}$ in the wavelength-intensity-matrix formalism\cite{alexandrowicz_helium_2007}. Full discussions of HeSE measurements and instrument design have been presented by Jardine \emph{et al.}\cite{Jardine2004} and Alexandrowicz \emph{et al.}\cite{alexandrowicz_helium_2007}. Specifics on HeSE phonon measurement and interpretation of phonon linewidths has been reported by Liu \emph{et al.}\cite{liu_application_2020,liu_distinguishing_2024}

\section{Results}

\subsection{Mechanical properties of a single layer of MoS\textsubscript{2}}

In 2015 Al Taleb \emph{et al.} introduced helium atom scattering (HAS) as a new technique to measure the bending rigidity in quasi-freestanding 2D materials \emph{via} the dispersion of the flexural mode\cite{AlTaleb2015HeliumFoil} using theory developed for graphene by Amorim \& Guinea \cite{Amorim2013FlexuralSubstrate}. The dispersion of the flexural mode contains a wealth of information on the mechanical properties of a quasi-freestanding monolayer and is defined as follows,
\begin{equation}
    \omega_{\mathrm{flex}}(q)=\sqrt{\frac{\kappa }{\rho_{2D} }q^4 +v_{\mathrm{flex}}^2 q^2 +\omega_0^2},
    \label{eqn:flexural}
\end{equation}
where $\kappa$ is the bending rigidity, $\rho_{2D}$ the areal density ($\rho_{2D}=\SI{3.06e-6}{\kilo\gram\per\metre\squared}$ for ML-MoS\textsubscript{2}), $v_{\mathrm{flex}}$ the sound velocity and $\omega_0$ the gap energy from the monolayer-substrate binding. The sound velocity also contains the out-of-plane shear stiffness, $c_{44}$, ($v_{\mathrm{flex}}=\sqrt{\nicefrac{c_{44}}{\rho_{2D}}}$), and the gap energy reveals the substrate coupling force, $g$, ($\omega_0 =\sqrt{\nicefrac{g}{\rho_{2D}}}$). The $q^2$ term corresponds to the inclusion of finite boundary conditions and has been omitted in previous experimental measurements of quasi-freestanding modes as it is usually overshadowed by the substrate binding energy\cite{Buchner2018BendingSilica}. We show that for materials with high defect densities, such as TMDs, the finite boundary condition must be included due to the presence of disorder.

\begin{figure*}[t]
    \centering
    \includegraphics[width=0.9\textwidth]{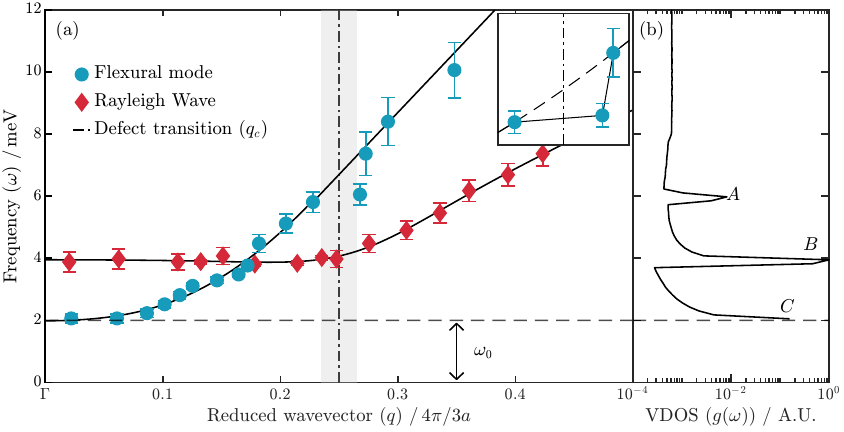}
    \caption{(a) Dispersion curves for flexural (circles) and hybrid Rayleigh Wave (hRW) (diamonds) modes of quasi-freestanding ML-MoS\textsubscript{2} along $\overrightarrow{\Gamma K}$. For $q\leq 0.25$ (vertical dashed line) the out-of-plane mode is well-described as a flexural mode from elastic membrane theory (Eqn. \ref{eqn:flexural}\cite{Amorim2013FlexuralSubstrate}) and the hRW is non-dispersive due to dominant defect scattering. For $q\geq 0.25$ both modes have positive linear dispersions. The regime change (vertical dashed line at $q_c=0.25\pm0.02$) occurs at a characteristic length $\lambda_c\sim1.90\pm\SI{0.15}{\nano\metre}$ that we attribute to the characteristic length between defects. Inset shows the creation of a Van Hove singularity (VHs) around $q_c$ in the flexural mode, corresponding to feature A in panel (b). (b) 1D Vibrational density of states (VDOS) along $\overrightarrow{\Gamma K}$ calculated from experimental dispersions reveals three VHs labeled A, B and C. VHs A and B arise from defect-phonon scattering, forming large enhancements in the VDOS away from any high symmetry points that will significantly affect thermal transport. VHs C is expected from elastic membrane theory due to the parabolic dispersion lineshape flattening near $\Gamma$. All data measured at $\SI{120}{\degreeCelsius}$. Fitting and calculation details for dispersions and VDOS are presented in the End Matter along with sample and measurement details.}
    \label{fig:phonons}
\end{figure*}

%\todo[inline,color=green!40]{Our work does beg a comparison with the predicted dispersion in truly freestanding monolayer MoS\textsubscript{2}, which is readily calculated (as opposed to MoS\textsubscript{2}-on-something)....}

Figure \ref{fig:phonons} shows the phonon dispersion along the $\overrightarrow{\Gamma K}$ direction measured with HeSE, two gapped low-energy modes were identified. Alongside the dispersion is plotted 1D the vibrational density of states (VDOS) along the $\overrightarrow{\Gamma K}$ direction calculated from fits to the experimental data. Appendix \ref{app:vdos} details the calculation of VDOS from the dispersion. The two modes are identified as the flexural mode, blue, and a hybrid Rayleigh Wave (hRW), red, as the two lowest energy modes present at the surface of a solid. 

Fitting the flexural mode to Eqn.\ref{eqn:flexural} yields $\kappa=\SI{12.0(2.5)}{\electronvolt}$, which agrees within uncertainty with the theoretical literature value $\kappa=\SI{9.61}{\milli\electronvolt}$ from Jiang et al.\cite{Jiang2013ElasticEffect}. At small wavevectors, the $q^{2}$ and $q^{4}$ contributions become comparable, making them weakly constrained by the data. As a result, the fitted pre-factors, and derived quantities such as the bending rigidity ($\kappa$) and sound velocity ($v$), carry correspondingly larger uncertainties.

%Fitting a dispersion containing both $q^{2}$ and $q^{4}$ terms is \hl{This is a bit negative, can we be less negative here?} intrinsically ill-conditioned at small $q$, where the two contributions become nearly degenerate, leading to correspondingly large uncertainties on the fitted coefficients. Therefore, values calculated from these pre-factors, such as bending rigidity ($\kappa$) and sound velocity ($v$), will also have relatively large uncertainties.

Although the flexural mode is out-of-plane acoustic (ZA), it has non-zero energy at the high symmetry point $\Gamma$ due to substrate coupling, gaining a gap frequency $\omega_0=\SI{1.99(0.16)}{\milli\electronvolt}$. Using the known areal density of ML-MoS\textsubscript{2}, $\rho_{2D}$, the gap frequency yields the substrate coupling strength, $g=\SI{2.79(45)e19}{\newton\per\metre\cubed}$ which agrees with Raman values for interlayer MoS\textsubscript{2}-MoS\textsubscript{2} ($\SI{2.82e19}{\newton\per\metre\cubed}$)\cite{Zhang2013Raman2}. The substrate coupling strength comes from approximation of the monolayer and substrate as a simple harmonic oscillator\cite{Amorim2013FlexuralSubstrate,Zhang2013Raman2}, thus confirming that the topmost layer of bulk MoS\textsubscript{2} vibrates as a quasi-freestanding monolayer. The surface will always support a Rayleigh wave, however, unlike the flexural mode which becomes gapped in the presence of substrate coupling the Rayleigh wave remains strictly gapless and near-linear in a defect-free material\cite{Amorim2013FlexuralSubstrate}. Furthermore, the Rayleigh wave is the only excitation expected with lower energy than the flexural mode at small $q$. We therefore assign the mode with frequency $\omega(\Gamma)=\SI{3.95}{\milli\electronvolt}$ in Figure \ref{fig:phonons} (red diamonds) to the hybrid Rayleigh Wave (hRW) and discuss the defect-induced frequency gap for $q<q_c$ in section \ref{sec:defect_scattering}.

The bending rigidity relates to the 3D Young's modulus, $Y$, by
\begin{equation}
    \kappa=\frac{Yh^3}{12(1-v^2)}
    \label{eqn:young_mod}
\end{equation}
where $v$ is Poisson's ratio and $h$ is the monolayer's thickness, which we take as the S-S distance in the c-axis. Poisson's ratio has been reported as $\SI{0.25}{}$ for ML-MoS\textsubscript{2} from atomic force microscopy and finite element calculations by Li \emph{et al.}\cite{Li2018MappingSimulation}, and $h=\SI{0.65}{\nano\metre}$. The 3D Young's modulus of ML-MoS\textsubscript{2} is therefore $Y=\SI{78.8(16.5)}{\giga\pascal}$, or equivalently $Y_{\mathrm{2D}}=\SI{51.2(10.8)}{\newton\per\metre}$ in two dimensions\footnote{Literature values of the 3D Young's modulus, $Y$ ($\SI{}{\giga\pascal}$), have been converted to the two-dimensional equivalent, $Y_{\mathrm{2D}}$ ($\SI{}{\newton\per\metre}$) by $Y_{\mathrm{2D}}=\nicefrac{Y}{h}$ where $h=\SI{0.65}{\nano\metre}$ for ML-MoS\textsubscript{2}.}. Our bending rigidity is significantly softer than theoretical and experimental results from Cooper \emph{et al.} where they calculated $Y_{\mathrm{2D,\,theory}}=\SI{130}{\newton\per\metre}$ using density function theory, and measured $Y=120\pm\SI{30}{\giga\pascal}$ using AFM indentation\cite{Cooper2013NonlinearDisulfide}. Bertolazzi \emph{et al.} also perform AFM indentation and report $Y_{\mathrm{2D,\,exp.}}=\SI{180(60)}{\newton\per\metre}$\cite{Bertolazzi2011Stretching2}. Our value is $\sim \times3$ smaller than literature values which can potentially be attributed to tension induced in the monolayer when suspended for AFM indentation measurements.

We find the velocity of sound in the flexural mode as $v_{\mathrm{flex}}=\SI{1381(340)}{\metre\per\second}$, which in turn gives an out-of-plane shear stiffness $c_{44}=\SI{9.0(45)}{\giga\pascal}$, agreeing well with literature values measured using a variety of techniques\cite{Zhang2013Raman2}. \todo[color=green!40]{no experimental values?}Literature calculations typically report far higher group velocities for the ZA mode, for example $<\SI{14}{\kilo\metre\per\second}$\cite{Rai2020ElectronicSOC}. A surprisingly low group velocity, in addition to a lack of true acoustic modes, will significantly affect in-plane thermal conductivity in mono- and few-layer MoS\textsubscript{2}. We discuss the low group velocity in more detail towards in section \ref{sec:thermal_conductivity}. 

\subsection{Defect-phonon scattering and Van Hove singularities}\label{sec:defect_scattering}

Figure \ref{fig:phonons} reveals a distinct crossover in the dispersion relations of both the flexural and hRW modes at reduced wavevector $q_c=\SI{0.25(02)}{}$, corresponding to a real-space characteristic length of $\lambda_c=\SI{1.90(1.6)}{\nano\metre}$, roughly six MoS\textsubscript{2} lattice parameters. Below $q_c$ the flexural mode follows the expected quadratic dispersion for an unbounded quasi-freestanding elastic membrane, including a linear term, usually overshadowed by the gap frequency, arising from finite boundary conditions\cite{AlTaleb2015HeliumFoil}. Simultaneously the hRW is non-dispersive. For $q>q_c$ both modes have positive linear dispersion. For the flexural mode this indicates a regime where finite boundary conditions are dominant, and the expected behavior for the hRW ($v_{\mathrm{hRW}}=\SI{2.15(20)}{\kilo\metre\per\second}$) from theory and experiment\cite{Amorim2013FlexuralSubstrate}. The simultaneous onset of disordered or atomistic vibrational behavior in the two modes at a wavevector corresponding to typical MoS\textsubscript{2} sulfur vacancy densities strongly suggests that these effects originate from defects.

The areal density of defects, $n_c$ is calculated by $n_c=\nicefrac{1}{\lambda_c}$, therefore $n_c=\SI{2.8(5)e13}{\per\centi\metre\squared}$, which corresponds to typical sulfur vacancy ($V_s$) densities in `pristine' MoS\textsubscript{2} crystals\cite{zhu_room-temperature_2023,HAMD_for_defects}. We therefore attribute the sharp change in dispersive behavior in both modes to defect scattering.

%\todo[inline,color=green!40]{The inclusion of the defect stuff and relating to our observations is really nice compared to earlier drafts. The distinction between the pre}

In a defect-free, free-standing monolayer, $\omega_{\mathrm{flex}}\propto q^2$ represents free bending without a restoring force. Substrate interaction and sulfur vacancies, however, introduce tension, producing a dispersion of the form Eqn. \ref{eqn:flexural}\cite{Mariani2008FlexuralGraphene,Los2009ScalingGraphene,AlTaleb2015HeliumFoil}. Atomistic simulations suggest that sulfur vacancies act as pinning centers that suppress out-of-plane motion and couple ZA displacements to in-plane shear\cite{Peng2016BeyondMoS2}, resulting in shortening of flexural phonon lifetimes. We present complementary measurements exploring the influence of defects on the linewidth of the flexural mode in section \ref{sec:linewidths_maintext}. The resulting crossover at $\lambda_c$ therefore marks the transition from a continuum bending regime to an atomistic, defect pinned regime.

The brief non-dispersive region, where $\partial\omega/\partial q\to 0$ in the flexural dispersion around $q_c=\SI{0.25(0.02)}{}$ (inset in Figure \ref{fig:phonons} (a)) creates an enhancement in the vibrational density of states (VDOS), and is a Van Hove singularity (VHs) (feature A in Figure \ref{fig:phonons} (b)). The emergence of a VHs at $\omega_\text{flex}(q_c)$ (feature A) can be explained by the formation of flexural standing waves between sulfur vacancies, where the wavelength of the flexural mode is approximately equal to the defect-defect characteristic distance, $L_d$. We find two further VHs, features B and C in Figure \ref{fig:phonons} (b), that lie near the high-symmetry point $\Gamma$. VHs C is expected from elastic membrane theory as the curvature of the flexural mode must approach 0 at $\Gamma$. VHs A and B, however, both emerge from strong defect-phonon scattering. Similar defect-induced flattening of the flexural branch has been predicted for graphene and other atomically thin membranes\cite{Mariani2008FlexuralGraphene,Zakharchenko2010AtomisticGraphene}, where pinned out-of-plane motion produces localized or resonant modes, significantly suppressing thermal conductivity. To our knowledge, a VHs in the VDOS has not previously been measured experimentally in a monolayer semiconductor away from the zone edge. Due to the lack of selection rules in HeSE, and approximating that MoS\textsubscript{2} is isotropic in-plane, our measurements of phonon dispersions capture all vibrations along $\overrightarrow{\Gamma K}$ and are thus representative of a slice through the full VDOS for the energy range measured, allowing unique insights into the atomistic mechanisms that determine thermal transport.

In contrast, the hRW is non-dispersive when $q<q_c$, where the phonon wavelength shortens and approaches the defect-defect characteristic distance. First-principles supercell calculations of MoS\textsubscript{2} containing sulfur vacancies reveal localized resonant phonon states that couple strongly to acoustic modes, producing the same flattening and reduced slope observed in this work\cite{Peng2016BeyondMoS2}. Our results demonstrate that atomic-scale disorder modifies vibrational dynamics long before the Brillouin-zone edge is reached. Notably, two of the observed VHs (features A and B in Figure \ref{fig:phonons}) arise from defects far away from BZ high-symmetry points. 

Ab initio calculations have shown that the defect scattering rate for acoustic phonons can rival 3-phonon scattering rates at room temperature\cite{Polanco2020Defect-limitedMoS2}, leading to a substantial suppression of lattice thermal conductivity $\kappa_L$\cite{Mahendran2024QuantitativeMonolayers}. Experimental measurements on suspended, defect-rich MoS\textsubscript{2} similarly report $\kappa_L$ values in the tens of $\SI{}{\watt\per\metre\per\kelvin}$, significantly below those of exfoliated `pristine' samples\cite{Yarali2017EffectsDeposition}. Sulfur vacancies have also been calculated to significantly broaden the linewidths of acoustic phonons rather than shift their energies \cite{Peng2016BeyondMoS2,Polanco2020Defect-limitedMoS2}. We present linewidth measurements of the hRW at $q<q_c$ that support these calculations qualitatively later in this Article.

These results establish a direct signature of vacancy-induced vibrational disorder in monolayer TMDs and identify the atomistic length scale at which the transition from continuum to defect-limited dynamics occurs. We discuss the effect of defect-induced vibrational disorder on thermal conductivity later in this Article.

\subsection{Phonon linewidths without local thermal effects}\label{sec:linewidths_maintext}

Accurate optical measurement of phonon linewidths is challenging in many 2D materials due to their poor thermal conductivity\cite{Zobeiri2020EffectInterference,Sokalski2022EffectsDisulfide} or weak Raman scattering cross-section\cite{Stenger2017LowCrystals}. The TMD family and hexagonal boron nitride (hBN) are of particular interest due to their technological relevance, which suffer from suppressed thermal conductivity and weak Raman cross-sections, respectively, in the few-layer regime specifically. Meanwhile the $\Delta\omega\sim\SI{1}{\micro\electronvolt}$ resolution\cite{Jardine2004} of HeSE allows the extraction of linewidths from inelastic features\cite{liu_distinguishing_2024,liu_experimental_2024}.

\begin{figure}[h]
    \centering
    \includegraphics[width=\linewidth]{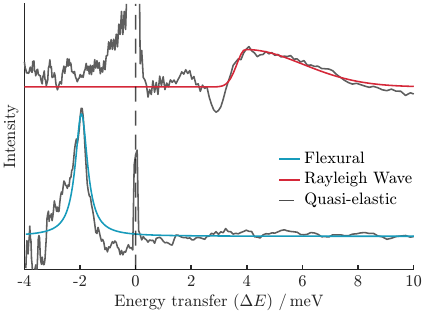}
    \caption{Example phonon linewidths of the flexural (blue) and hybridized Rayleigh Wave (hRW, red) near $\Gamma$ ($q=0.02$). The flexural mode is described by a Lorentzian linewidth $\gamma = \SI{0.54(5)}{\milli\electronvolt}$ ($\SI{4.36(41)}{\per\centi\metre}$). The hRW has an ill-defined linewidth as expected in a non-dispersive region ($q<q_c$). An asymmetric Gaussian is shown to guide the eye. We note that the flexural mode is always measured as an creation peak ($-\mathrm{ive}\,\Delta E$) whereas the hRW is annihilation-type. Data has been normalised and scaled for visual clarity.}
    \label{fig:lorentzian_fitting}
\end{figure}

%\hl{Re work the below without the comparison to the Raman data as they are measuring a different mode... Can cite Najmaei 2012 as evidence of the effect of lasers on the measured linewidth. Boukhicha et al. do have measurements of what they term the "compression mode" for bilayer systems, but not monolayer.}
In figure \ref{fig:lorentzian_fitting} we present an energy-domain HeSE spectrum where an inelastic (phonon) feature has been fit with a symmetric Lorentzian, with a polynomial baseline correction, whose full-width at half-maximum (FWHM) relates to a single phonon's linewidth, $\gamma$, by $\gamma=2\cdot\mathrm{FWHM}$ to find a linewidth for the flexural mode of $\gamma_{\mathrm{flex}}=\SI{0.54(5)}{\milli\electronvolt}$ ($\SI{4.36(41)}{\per\centi\metre}$). We do not fit a Voigt function because the instrument response function of the HeSE instrument is significantly narrower than the observed phonon linewidths\cite{Jardine2004}. Details on the measurement of phonon linewidth using HeSE has been previously published\cite{liu_experimental_2024,liu_distinguishing_2024}. To our knowledge these are the first measurements of the linewidth of an acoustic mode in monolayer MoS\textsubscript{2}. Although some optical methods can access the energy range presented in this Article, the acoustic modes' symmetries prevent measurement. For linewidth measurements for other modes in few-layer MoS\textsubscript{2} Najmaei \emph{et al.} have demonstrated a strong dependence of the mode's linewidth on fluence, attributing rapid increases in linewidth (and therefore decrease in lifetime) with thermal effects\cite{Najmaei2012}, for example for the $A_{1g}$ mode they report that decreasing laser fluence $20\rightarrow\SI{2}{\milli\watt}$ decreases the mode's linewidth significantly, $7.1\rightarrow\SI{5.4}{\per\centi\metre}$. HeSE measurements do not produce any local heating and therefore cannot increase the rate of phonon-phonon scattering. The measurement of phonon linewidths in 2D materials, free from thermal effects and selection rules, is a notable advantage of HAS techniques.
%although results are present for bilayer\cite{boukhicha_anharmonic_2013} where $\SI{6.7}{\per\centi\metre}$ was reported and for multilayer (3-5 layers) $\approx\SI{4}{\per\centi\metre}$.

The linewidth relates to lifetime, $\tau$, by
\begin{equation}
    \gamma = \frac{\hbar}{\tau}.
    \label{eqn:linewidth}
\end{equation}
We therefore report the lifetime of the flexural mode as $\tau_{\mathrm{flex}}=\SI{1.21(11)}{\pico\second}$. The strong defect scattering experienced by the in-plane mode for $q<q_c$ makes it difficult to define a linewidth. The typical spectrum for the in-plane mode, shown in Figure \ref{fig:lorentzian_fitting}, is very broad, making linewidth measurement, and definition, challenging. This is expected if the mode experiences significant structural disorder on the appropriate length-scale, akin to `boson peaks' that are  observed in glassy or liquid systems\cite{Gonzalez-Jimenez2023UnderstandingGlasses}.

Approximating the velocity of the flexural mode using its linear component, $v_{\mathrm{flex}}$, allows us to also calculate the mean-free path, $\Lambda$, \emph{via} the following relation,
\begin{align}
    \Lambda&=\tau\cdot v(q),%\\
     %v(q) =\frac{\partial \omega(q)}{\partial q} &= \frac{4(\nicefrac{\kappa}{\rho_{2D}})q^3+2(v_{\mathrm{flex}}^2)q}{2\sqrt{\nicefrac{\kappa}{\rho_{2D}}q^4+v_{\mathrm{flex}}^2q^2+\omega_0^2}},
     \label{eqn:mean_free_path}
\end{align}
thus returning $\Lambda_{\mathrm{flex}}=\SI{1.68(44)}{\nano\metre}$. Treating $\Lambda$ as the characteristic distance between defects we can calculate a defect density $\SI{3.5(1.9)e13}{\per\centi\metre\squared}$, agrees with the dispersion-derived defect density $\SI{2.8(5)e13}{\per\centi\metre\squared}$ within error bounds. We therefore suggest that the mean free path, and the critical wavevector ($q_c$) in Figure \ref{fig:phonons}, correspond to the defect-defect characteristic distance. The dispersion-derived value is significantly more precise because of the poor parameter conditioning affecting affecting the fitting of the $q<q_c$ region of the flexural mode, in turn creating a large uncertainty in $v_{\mathrm{flex}}$. 

The dispersion- and linewidth-derived defect densities agree to within their uncertainties and lie within typical values for untreated MoS\textsubscript{2} crystals, usually given as $\sim\SI{1e13}{\per\centi\metre\squared}$. Liu \emph{et al.} have shown that all terms contributing to phonon linewidth, electron-phonon, phonon-phonon and defect-phonon scattering, can be characterised with HeSE\cite{liu_experimental_2024}. A series of temperature dependent linewidth measurement would be required to retrieve each term's contribution to the linewidth.

\subsection{The microscopic origins of poor thermal conductivity in few-layer TMDs}
\label{sec:thermal_conductivity}

Many theoretical efforts have sought to predict the lattice thermal conductivity, $\kappa_L$ of mono- and few-layer MoS\textsubscript{2}, with reported values spanning over an order of magnitude ($83-\SI{841}{\watt\per\metre\per\kelvin}$) \cite{Wu2021,Ding2015,Peng2016,Li2013,Cai2014,Xu2016}. This wide variation extends across the transition-metal dichalcogenide (TMD) family and highlights the limitations of current models in describing anharmonic and defect-mediated transport in two-dimensional systems. In contrast, experimentally measured values for suspended MoS\textsubscript{2} samples are far more consistent, typically ranging between $34-\SI{52}{\watt\per\metre\per\kelvin}$, despite differences in measurement techniques such as micro-Raman spectroscopy \cite{Yan2014,Sahoo2013} and the bridge method \cite{Jo2014}. %These experimental conductivities are systematically lower than the majority of theoretical predictions, suggesting that conventional treatments of phonon scattering underestimate intrinsic damping in few-layer systems.

However, recent calculations demonstrated that four-phonon interactions are key to predicted thermal conductivity of monolayer MoS\textsubscript{2}, with their inclusion lowering $\kappa_L$ from $\kappa_{\mathrm{L,\,3-ph.}}=\SI{116}{\watt\per\metre\per\kelvin}$ to $\kappa_{\mathrm{L,\,4-ph.}}=\SI{25}{\watt\per\metre\per\kelvin}$ suggesting that higher-order anharmonicity is the dominant scattering mechanism in this regime\cite{Chaudhuri2024UnderstandingMoS2}.

%\todo[inline,color=green!40]{This next paragraph needs to be argued more clearly.}
%\todo[inline,color=green!40]{To what extent are the arguments here novel, and to what extent are they based on the theoretical work "Beyond Perturbation: Rle of Vacancy-Induced..."?}

%The surprisingly low, and even near-zero, acoustic phonon group velocities measured in this work combined with defect-induced vibrational disorder render many expected decay channels kinematically forbidden. The resulting suppression of 3-phonon Normal processes leaves 3-phonon Umklapp (thermally resistive) and 4-phonon Normal processes as the dominant heat carriers. The Van Hove-like enhancement in the VDOS, \todo[color=green!40]{It isn't shown in the figure, but inferred in the discussion}shown in Figure \ref{fig:phonons}, also serves as a `trap' where phonons accumulate and scatter repeatedly between defect sites rather than propagating ($\nicefrac{\partial \omega}{\partial q}|_{q\approx q_c}\rightarrow 0$). 

%Small group velocities, shortened acoustic lifetimes and local enhancements in VDOS all contribute to phonon scattering

The observed features, including low acoustic velocities, suppressed dispersions, shortened acoustic lifetimes, and the formation of an enhancement in the VDOS near $q_c$, arise from microscopic structure whose interactions with phonons are amplified in the few-layer limit. Defects and interlayer coupling strength are computationally expensive to capture in first-principles models, sometimes prohibitively so. By experimentally resolving these effects, HeSE provides direct access to the vibrational mechanisms that control thermal transport in two-dimensional semiconductors. In addition to explaining the anomalously low conductivity of MoS\textsubscript{2}, this approach establishes a framework for engineering phonon behavior through controlled modification of microscopic disorder, dopants, or heterostructure interfaces.

\section{Conclusion}

We have presented the first measurements of phonon dispersion in a monolayer semiconductor (MoS\textsubscript{2}). We use helium-3 spin-echo spectroscopy to characterize the flexural and hybridized Rayleigh Wave (hRW) modes and show that the topmost layer of bulk MoS\textsubscript{2} has the vibrational properties of a quasi-freestanding monolayer, allowing for facile measurement without sample suspension or preparation.

The dispersion of the flexural mode allows us to extract key mechanical properties of ML-MoS\textsubscript{2} and identify a critical wavevector, $q_c$, at which continuum descriptions of acoustic phonons in 2D materials break down, and atomistic effects become dominant. We attribute this wavevector to the characteristic distance between defects ($q_c\rightarrow L_D\approx\SI{1.9}{\nano\metre}$), creating standing flexural waves that lie between sulfur vacancies. The standing waves cause a local flattening in the flexural dispersion at $q_c$, resulting in a Van Hove singularity in the vibrational density of states.

The hybridized Rayleigh Wave undergoes a sharp regime change at the same critical wavevector corresponding to defect-defect characteristic distance, from non-dispersive to linear. For $q<q_c$ the hRW is vibrationally disordered (near-zero group velocity and an ill-defined, asymmetric linewidth), due to strong phonon-defect scattering.

Linewidth measurements of the flexural mode near $\Gamma$ return a lifetime $\tau_{\mathrm{flex}}=\SI{1.21(11)}{\pico\second}$. The flexural mode linewidth and group velocity (in the continuum limit) allow us to calculate a mean-free path $\Lambda_{\mathrm{flex}}=\SI{1.68(44)}{\nano\metre}$. The mean-free path gives an areal defect density of $\SI{3.5(1.9)e13}{\per\centi\metre\squared}$ agrees to within error bounds with our defect-defect characteristic distances extracted from the flexural and hybridized Rayleigh Wave dispersion curves ($\Lambda_{\mathrm{flex}}=\SI{1.9(2)}{\nano\metre}$), demonstrating two independent methods of extracting defect density in a monolayer material.

Our observation of strongly suppressed acoustic group velocities and Van Hove singularities far from the BZ edges provides experimental support for recent calculations predicting that four-phonon scattering processes play a dominant role in thermal transport in mono- and few-layer MoS\textsubscript{2}\cite{Chaudhuri2024UnderstandingMoS2}.

Helium atom scattering provides direct experimental access to phonon dispersions and linewidths in unprepared monolayer materials, enabling vibrational dynamics to be resolved in the presence of atomic-scale disorder. By revealing how disorder modifies low-energy acoustic phonons, this work establishes a microscopic framework for understanding vibrations, and therefore thermal transport, in two-dimensional materials. More broadly, these results clarify the role of disorder in limiting thermal transport in atomically thin semiconductors and demonstrate how momentum-resolved phonon measurements can uncover the breakdown of continuum descriptions in low-dimensional systems.

\subsection*{Data availability}
The data underlying all figures in the main text and Supplementary Information will be made publicly available from the University of Cambridge repository upon publication. All code used in this work is available from the corresponding authors upon reasonable request. 

\begin{comment}
    
\begin{acknowledgments}
Thank jack.
Thank you jack.
\end{acknowledgments}
\end{comment}

\clearpage

\appendix

\renewcommand{\thefigure}{A\arabic{figure}}
\setcounter{figure}{0}

\renewcommand{\thetable}{A\Roman{table}}
\setcounter{table}{0}

\renewcommand{\theequation}{A\arabic{equation}}
\setcounter{equation}{0}

\section{Phonon dispersion fitting}\label{sec:fitting_dispersions}

Model used to fit the flexural phonon mode in Figure \ref{fig:phonons} as it transitions from infinite ($q<0.25$) to finite ($q>0.25$) boundary condition elastic,
\begin{equation}
\omega_{\mathrm{flex}}(q)=T\cdot\underbrace{\vphantom{\left(\frac{1}{1}\right)}\sqrt{\frac{\kappa }{\rho_{2D} }q^4 +v_0^2 q^2 +\omega_0^2}}_{\substack{\text{Continuum limit}\\(q\lesssim0.25)}} + (1-T)\cdot  \underbrace{\vphantom{\left(\frac{1}{1}\right)}(v_1q^2+\omega_1)}_{\substack{\text{Defect pinning}\\\left(q\gtrsim0.25\right)}},
\label{eqn:flex_fitting}
\end{equation}
where $T$ is a transition function.

Model used to fit the hybridized Rayleigh Wave phonon mode in Figure \ref{fig:phonons} as it transitions from a non-dispersive, vibrationally disordered regime ($q\leq0.25$) to positive, linear dispersion,

\begin{equation}
\omega_{\mathrm{hRW}}(q)=T\cdot \underbrace{(v_0 q+\omega_0)}_{q\lesssim0.25} + (1-T)\cdot \underbrace{(v_1 q+\omega_1)}_{q\gtrsim0.25}.
\label{eqn:inplane_fitting}
\end{equation}

The transition function, $T$, is defined as,

\begin{equation}
T=1-\nicefrac{1}{2}\tanh{\left(\frac{q-q_c}{w}\right)},
\end{equation}
where the regime change occurs at $q_c$ with width $w$. $T$ is smooth for all real values.

\section{Calculation of the vibrational density of states (VDOS)}
\label{app:vdos}

The vibrational density of states (VDOS) (Figure \ref{fig:phonons} (b)) was obtained from the fitted phonon dispersions (Figure \ref{fig:phonons} (a)) using a Jacobian-weighted histogram method. For each phonon mode, the dispersion $\omega(q)$ was sampled uniformly in wavevector over the experimentally measured range, and the corresponding group velocity $v(q)=\partial\omega/\partial q$ was evaluated numerically. The VDOS was then constructed as
\begin{equation}
    g(\omega) \propto \sum_q \frac{1}{|v(q)|},
\end{equation}
which follows from transforming variables from $q$ to $\omega$. The sampled frequencies were binned, with each point contributing a weight $1/|\partial\omega/\partial q|$ to its corresponding frequency bin, and the resulting distribution was normalised to unit area.

To incorporate the defect-induced regime change of the flexural mode, we replaced the fitted flexural dispersion with linear interpolation between the experimental points surrounding $q_c$. The interpolation is shown in Figure \ref{fig:phonons} (a) inset. A narrow Savitsky-Golay filter was applied to this region to smooth discontinuities arising from the interpolation. The modified dispersion was then used in place of the analytic model (Eqn. \ref{eqn:flexural}) when evaluating both $\omega(q)$ and $\partial\omega/\partial q$.

\begin{comment}
\section{Phonon linewidths}\label{sec:fitting_linewidths}

The linewidths of inelastic features in HeSE can be characterized using Lorentzian functions and represent lifetimes\cite{liu_distinguishing_2024,liu_experimental_2024}. In figure \ref{fig:lorentzian_fitting} we perform a polynomial baseline correction and fit a single, symmetric Lorentzian whose full-width at half-maximum (FWHM) relates to a single phonon's linewidth, $\gamma$, by $\gamma=2\cdot\mathrm{FWHM}$. The lifetime, $\tau$, can then be calculated as $\tau=\nicefrac{\hbar}{\gamma}$ where $\hbar$ is the reduced Planck constant\cite{boyao_thesis}. We do not fit a Voigt function because the instrument response function of the HeSE instrument is significantly narrower than the observed phonon linewidths, with energy and momentum resolution $\Delta \omega\approx\SI{1}{\micro\electronvolt}$ ($\approx\SI{0.008}{\per\centi\metre}$) and $\Delta K\approx\SI{0.01}{\per\angstrom}$, respectively\cite{Jardine2004}.

\end{comment}

%\nocite{*}
\newpage
\bibliography{mos2_phonons}% Produces the bibliography via BibTeX.

\end{document}